\documentclass[aps,pra,twocolumn,superscriptaddress,showpacs]{revtex4-2}

\usepackage[utf8]{inputenc}
\usepackage{multirow}
\usepackage{array}
\usepackage{amsmath}
\usepackage{graphicx}
\usepackage{epstopdf}
\usepackage[export]{adjustbox}
\usepackage{color}
\usepackage{xcolor}
\usepackage{amsfonts}
\usepackage{subfigure}
\usepackage{amssymb}
\usepackage{graphicx}
\usepackage{makecell}
\usepackage[colorlinks,citecolor=blue]{hyperref}
\usepackage{changes}
\def\bra#1{\mathinner{\langle{#1}|}}
\def\ket#1{\mathinner{|{#1}\rangle}}

\newcommand{\onecolm}{
  \end{multicols}
  \vspace{-3.5ex}
  \noindent\rule{0.5\textwidth}{0.1ex}\rule{0.1ex}{2ex}\hfill
}
\newcommand{\twocolm}{
  \hfill\raisebox{-1.9ex}{\rule{0.1ex}{2ex}}\rule{0.5\textwidth}{0.1ex}
  \vspace{-4ex}
  \begin{multicols}{2}
}

\begin{document}

\title{Exact mobility edges in high-dimensional quasiperiodic networks}
\date{\today}

\author{Jia-Hui Zhang}

\affiliation {Key Laboratory of Atomic and Subatomic Structure and Quantum Control (Ministry of Education), Guangdong Basic Research Center of Excellence for Structure and Fundamental Interactions of Matter, School of Physics, South China Normal University, Guangzhou 510006, China}
\affiliation {Guangdong Provincial Key Laboratory of Quantum Engineering and Quantum Materials, Guangdong-Hong Kong Joint Laboratory of Quantum Matter, Frontier Research Institute for Physics, South China Normal University, Guangzhou 510006, China}

\author{Shan-Zhong Li}

\affiliation {Key Laboratory of Atomic and Subatomic Structure and Quantum Control (Ministry of Education), Guangdong Basic Research Center of Excellence for Structure and Fundamental Interactions of Matter, School of Physics, South China Normal University, Guangzhou 510006, China}
\affiliation {Guangdong Provincial Key Laboratory of Quantum Engineering and Quantum Materials, Guangdong-Hong Kong Joint Laboratory of Quantum Matter, Frontier Research Institute for Physics, South China Normal University, Guangzhou 510006, China}

\author{Zhi Li}
\email{lizphys@m.scnu.edu.cn}
\affiliation {Key Laboratory of Atomic and Subatomic Structure and Quantum Control (Ministry of Education), Guangdong Basic Research Center of Excellence for Structure and Fundamental Interactions of Matter, School of Physics, South China Normal University, Guangzhou 510006, China}
\affiliation {Guangdong Provincial Key Laboratory of Quantum Engineering and Quantum Materials, Guangdong-Hong Kong Joint Laboratory of Quantum Matter, Frontier Research Institute for Physics, South China Normal University, Guangzhou 510006, China}

\begin{abstract}
Inspired by the one-dimensional hidden self-duality [H.-T. Hu et al., Phys. Rev. Lett. 134, 246301 (2025)], we establish a generalized theoretical framework for a specific class of high-dimensional quasiperiodic network models. Specifically, as long as the periodic network acts as bond-decorating links and the quasiperiodic potentials are spatially separable, the system can be mapped onto an effective high-dimensional generalized Aubry-André equation by integrating out the periodic sites. This mapping restores the hidden self-duality symmetry in these complex models, thereby yielding exact mobility edges. Taking two and three-dimensional models as examples, we demonstrate the consistency between the theoretical prediction and the numerical results. This paper establishes a clear paradigm for mapping high-dimensional quasiperiodic networks to high-dimensional Aubry-André models, providing a systematic pathway for the construction and exploration of the exact mobility edge in high-dimensional systems.
\end{abstract}

\maketitle

\section{INTRODUCTION}
In disordered systems, the quantum interference effect causes the properties of the eigenstates to be energy-dependent, resulting in extended-localized phase transition. The boundary between these phases is known as the mobility edge—a core concept in Anderson localization theory that provides an important basis for understanding metal-insulator transitions and electron transport phenomena \cite{Anderson1958,Billy2008,Lee1985,Roati2008,Mott1987}. On the other hand, the single-parameter scaling theory indicates that the Anderson transition in random disordered systems exhibits strong dimension-dependence \cite{Thouless1974,MacKinnon1981,Sarker1981,Abrahams1979}. In the three-dimensional case, a mobility edge can emerge \cite{Evers2008}, but this is not the case in low-dimensional (one- and two-dimensional) systems. Specifically, even the slightest degree of random disorder will cause all the eigenstates of the low-dimensional system to enter the localized state \cite{Lee1981,Schwartz2007,Kramer1993}.

Quasiperiodic systems represent an intermediate regime between periodic systems and random disordered systems. In the low-dimensional case, they also support the emergence of localization transitions and the mobility edge \cite{SergeAubry1980,DasSarma1988,Danieli2015,Deng2019,Li2017,Izrailev2012,Yuan2000,Biddle2009,Yao2019,Li2016,Wang2022,Wang2023a,Hou2026}. For such systems, there exist a series of analytical methods, such as generalized self-duality \cite{SergeAubry1980,Wang2021,Biddle2010,Ganeshan2015}, Avila's global theory \cite{Avila2015,Wang2020}, and renormalization group \cite{Goncalves2023a,Goncalves2023}. Based on these methods, many models with exact mobility edges have been constructed. 

However, extending these analytical results to higher dimensions remains a formidable challenge. High-dimensional lattices typically exhibit more complex geometric connectivity patterns, which break the symmetry required for duality transformations. Furthermore, standard theoretical tools based on Avila's global theory cannot be further extended to higher dimensions. These limitations also explain why numerical studies dominate the landscape of two- and three-dimensional model investigations \cite{Strkalj2022,Yang2024,Szabo2020,Bordia2017,Huang2019,Gautier2021,Roche1997,Wang2023b,Li2025a}. While numerical methods can unveil rich phase diagrams, they are constrained by computationally accessible system sizes, and the absence of a unified analytical framework poses a fundamental barrier to obtaining high-dimensional models with exact mobility edges. Recently, some progress has been made in the study of exact solutions for two-dimensional quasiperiodic systems \cite{Duncan2024,Rossignolo2019}, such as a vertex-decorated Lieb lattice \cite{Wang2023,Jiang2024} and a two-dimensional non-Hermitian lattice \cite{Xu2021}. However, in the latter model, the dual space effectively decomposes into multiple uncoupled one-dimensional chains, thereby bypassing true high-dimensional spatial interference. While these models offer valuable insights into high-dimensional mobility edges, exactly solvable two- and three-dimensional quasiperiodic models remain scarce. Therefore, there is a fundamental need to develop systematic methods for constructing high-dimensional quasiperiodic systems with analytically solvable mobility edges.
        
Furthermore, the discovery of the hidden self-duality mechanism in the one-dimensional quasiperiodic network model \cite{Hu2025} has provided a different perspective for constructing a generalized theoretical framework applicable to higher dimensions. Based on this, one can design periodic points that can be integrated out and map the Hamiltonian into an energy-dependent effective tight-binding model. While the original model lacks explicit self-duality due to its complex geometric connectivity, a hidden self-dual structure emerges once the system is projected onto the quasiperiodic site subspace by integrating out the periodic sites. Within this specific subspace, the effective Schrödinger equation strictly maps onto a generalized Aubry-André (AA) model. Because this effective model restores the generalized duality transformation symmetry between real and momentum spaces, the corresponding self-dual condition directly yields the exact analytical solution for the mobility edges of the original model.

In this study, we extend the concept of hidden self-duality from one-dimensional quasiperiodic networks to high-dimensional systems. We consider a specific class of $n$-dimensional models composed of periodically arranged lattice points and quasiperiodic primary sites. By integrating out the periodic sites, we map the original Hamiltonian to a generalized effective AA model defined on the quasiperiodic lattices. Notably, we demonstrate that previously solved structures, such as the Lieb lattice \cite{Wang2023}, are special cases of the theory presented in this paper. Using this principle, we design and solve various two- and three-dimensional quasiperiodic network models. The results show that the mobility edges of these systems are all determined by a unified self-duality condition and exhibit robustness with respect to the spatial dimension. These findings provide a theoretical basis for controlling wave localization in complex high-dimensional lattices.

\section{Hidden Self-Duality and Exact Mapping}
\begin{figure}[t] 
    \centering
    \includegraphics[width=\linewidth]{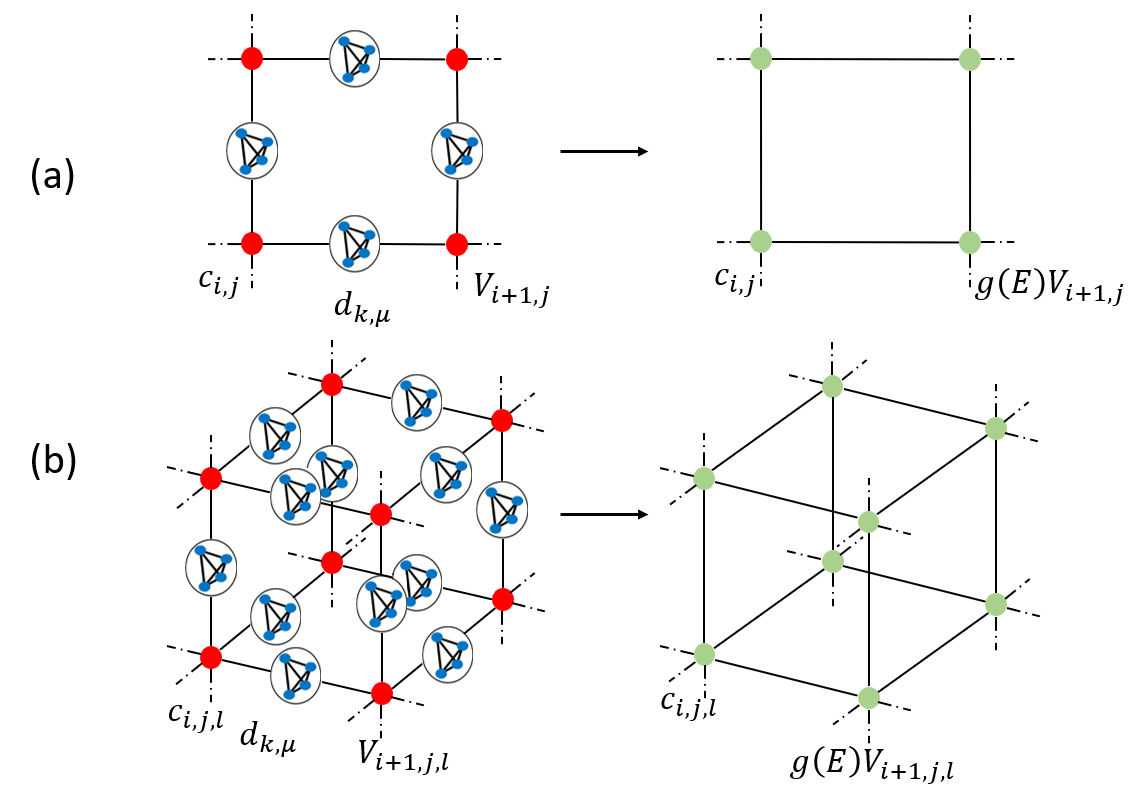}
    \caption{The generalized (a) two-dimensional and (b) three-dimensional quasiperiodic network model constituted by periodic network sites between quasiperiodic sites and their corresponding reduced effective models. Here, $V_{i+1,j}(V_{i+1,j,l})$ and $g(E)V_{i+1,j}(g(E)V_{i+1,j,l})$ denote the potentials in 2D (3D) before and after integrating out the periodic network sites. Here, $d_{k,\mu}$ denotes the $k$-th intermediate periodic site on the specific bond along the $\mu$-direction (e.g., $\mu = x, y$). The red, blue, and green solid points represent the lattice sites hosting the quasiperiodic, the periodic, and the effective energy-dependent potential, respectively. }
    \label{fig.1}
\end{figure}

We consider a specific subset of generalized single-particle quasiperiodic network models defined on an $n$-dimensional ($n \ge 1$) hypercubic lattice. Our analytical framework rigorously applies to models satisfying two structural conditions: (1) periodic network acts exclusively as bond-decorating links between nearest-neighbor quasiperiodic sites, without cross-dimensional coupling, and (2) the quasiperiodic potential is spatially separable, as illustrated in Fig.~\ref{fig.1}. These two prerequisites fundamentally guarantee that the system can be exactly mapped onto a solvable high-dimensional AA model after integrating out the periodic degrees of freedom. The total Hamiltonian of the system is given by
\begin{equation}
    H=H_q+H_p+H_c,
\end{equation}
where $H_q$ denotes the Hamiltonian of the primary quasiperiodic lattice. It incorporates an on-site separable quasiperiodic potential $V_{\mathbf{r}} = \sum_{\mu=1}^n V \cos(2\pi\alpha_\mu r_\mu + \phi_\mu)$, where $\mathbf{r}=(r_1,r_2,...,r_n)$ denotes the $n$-dimensional spatial coordinates, $V$ is the quasiperiodic potential amplitude, $\alpha_\mu$ is a multidimensional irrational number along the $\mu$-th spatial direction, whose incommensurability prevents the potential from degenerating into a periodic superlattice governed by Bloch's theorem \cite{SergeAubry1980}, and the phase shift is set to $\phi_\mu=0$. $H_p$ describes the Hamiltonian for periodic network connecting to the quasiperiodic sites, termed the network, and $H_c$ characterizes the coupling between the quasiperiodic sites and the periodic network. Denoting the eigenvector as $|\psi\rangle=\left(\psi_q,\;\psi_p\right)^T$, the Hamiltonian can be expressed in a partitioned matrix form
\begin{equation}
    \begin{pmatrix}
    H_q & H_c \\
    H_c^\dagger & H_p
    \end{pmatrix}
    \begin{pmatrix}
    \psi_q \\
    \psi_p
    \end{pmatrix}
    = E
    \begin{pmatrix}
    \psi_q \\
    \psi_p
    \end{pmatrix},
\end{equation}
\begin{table*}[t] 
    \centering 
    \caption{Summary of three representative 2D lattice models and their physical properties. The columns correspond to model I (extended Lieb), model II (rhombic-decorated), and model III (densely-connected), respectively. The top panels illustrate the geometric structure of each model, where large red points represent the quasiperiodic primary sites $V_{i,j}$ and small blue points denote the periodic network sites connecting them, whose bare onsite energies are denoted by $U_k$. The bottom two rows list the corresponding potential renormalization factor $g(E)$ and the effective energy characteristic function $f(E)$, with a detailed derivation provided in the Appendix \ref{Appendix A}. }
    \label{Table I}
    \begin{tabular}{|c|c|c|c|}
        \hline
        \hline
         & \rule{0pt}{22pt} \begin{tabular}{@{}c@{}}\textbf{Model I} \\ \textbf{(Extended Lieb)}\end{tabular} & \begin{tabular}{@{}c@{}}\textbf{Model II} \\ \textbf{(Rhombic decorated)}\end{tabular} & \begin{tabular}{@{}c@{}}\textbf{Model III} \\ \textbf{(Densely connected)}\end{tabular} \\ 
        Structure & 
        \begin{minipage}{0.25\textwidth} \centering \vspace{1pt} \includegraphics[width=\linewidth]{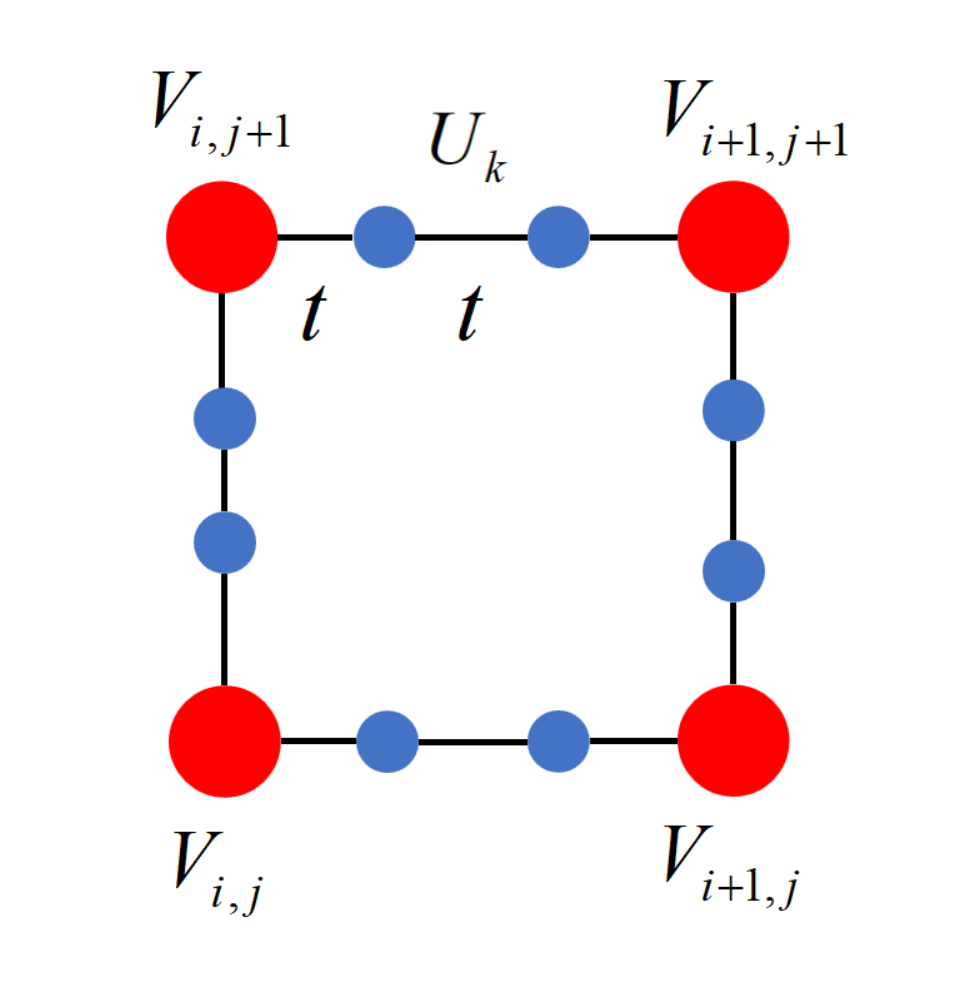} \vspace{1pt} \end{minipage} & 
        \begin{minipage}{0.25\textwidth} \centering \vspace{1pt} \includegraphics[width=\linewidth]{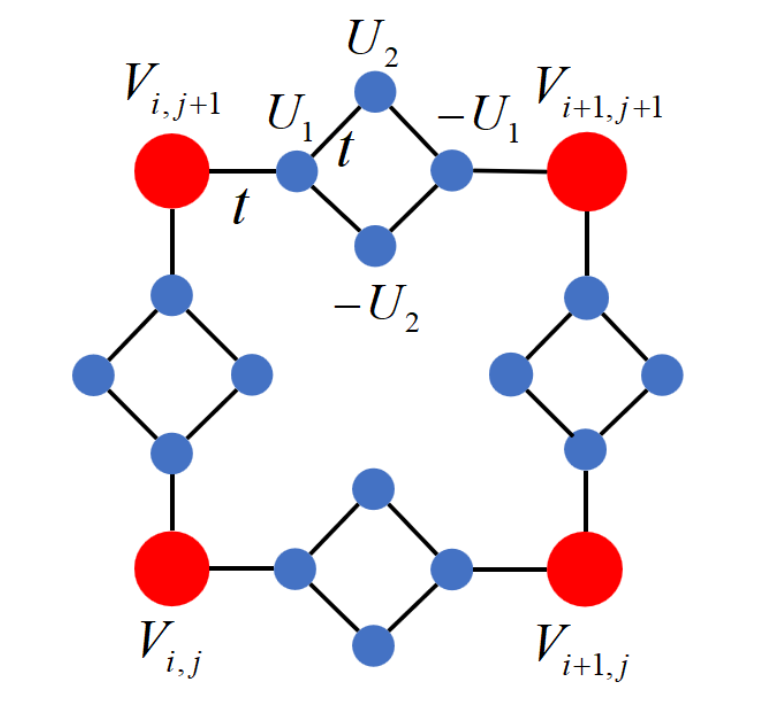} \vspace{1pt} \end{minipage} & 
        \begin{minipage}{0.25\textwidth} \centering \vspace{1pt} \includegraphics[width=\linewidth]{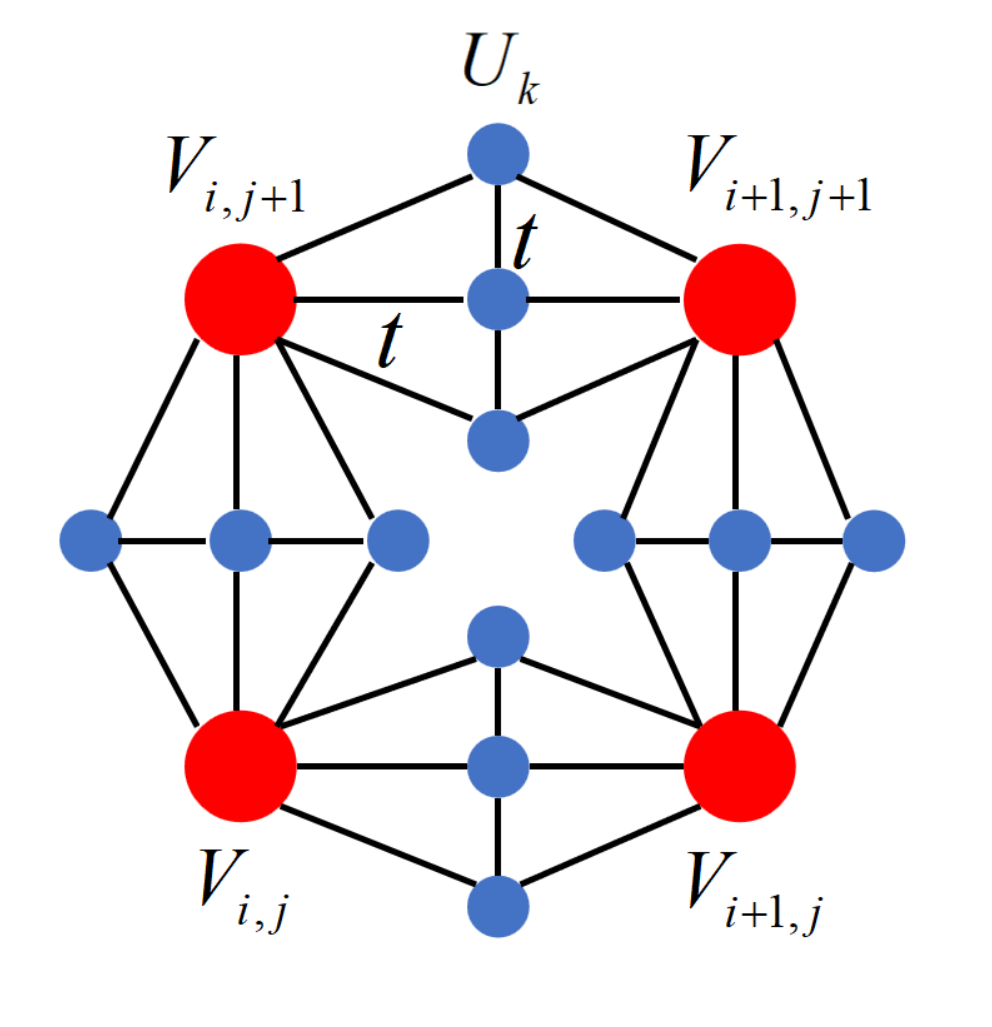} \vspace{1pt} \end{minipage} \\ \hline
        
        $g(E)$ & 
        \rule{0pt} {15pt} $t^2-E^2$ & 
        \rule{0pt}{15pt} $\displaystyle \frac{E^4-E^2(U_1^2+U_2^2+4t^2)+U_1^2U_2^2}{2E}$& 
        \rule{0pt}{15pt} $\displaystyle \frac{E^2-2t^2}{3E-4t}$ \\[10pt]
        
        $f(E)$ & 
        \rule{0pt}{15pt} $E(5t^2-E^2)$ & 
        \rule{0pt}{15pt} $\displaystyle \frac{E^4-E^2(U_1^2+U_2^2+8t^2)+U_1^2U_2^2+4U_2^2t^2+8t^4}{2}$ & 
        \rule{0pt}{15pt} $\displaystyle \frac{E^3-14Et^2+16t^3}{3E-4t}$ \\ \hline
        \hline
    \end{tabular}
    
\end{table*}
where $\psi_q$ and $\psi_p$ denote the wave functions of the quasiperiodic and periodic sites, respectively, and $E$ is the eigenenergy. The periodic wave function $\psi_p$ can be expressed as $\psi_p=(E-H_p)^{-1}H_c^\dagger \psi_q$. Substituting this expression back into the equation for the quasiperiodic sites yields an energy-dependent effective equation $H_{eff}(E)\psi_q=E\psi_q$ wherein the energy-dependent effective Hamiltonian is defined as  
\begin{equation}
    H_{eff}(E)=H_q+H_c(E-H_p)^{-1}H_c^\dagger.
\end{equation}
Projecting this effective Hamiltonian onto the basis of quasiperiodic sites $\ket{\mathbf{r}}$, the first part contributes solely to the diagonal terms due to the absence of direct inter-site hopping in the primary lattice, $\bra{\mathbf{r}}H_q\ket{\mathbf{r}'}=V_{\mathbf{r}}\delta_{\mathbf{r},\mathbf{r}'}$. The second term can be separated into a diagonal term $W_{diag}(E) = \bra{\mathbf{r}} H_c (E-H_p)^{-1} H_c^\dagger \ket{\mathbf{r}}$ and an off-diagonal term $W_{off}(E) = \bra{\mathbf{r}} H_c (E-H_p)^{-1} H_c^\dagger \ket{\mathbf{r} \pm \hat{e}_\mu}$, where $\hat{e}_\mu$ is the unit vector in the $\mu$-direction. Since the periodic links lack direct cross-couplings, both $H_p$ and $(E-H_p)^{-1}$ are strictly block-diagonal. This geometric isolation rigorously restricts $W_{off}$ to nearest-neighbor hopping, preventing any higher-order couplings. Substituting these matrix elements into the effective eigenvalue equation $\sum_{\mathbf{r}'} \bra{\mathbf{r}}H_{eff}(E) \ket{\mathbf{r}'} \psi_{\mathbf{r}'}=E\psi_{\mathbf{r}}$, where $\psi_\mathbf{r}=\langle \mathbf{r} \vert{} \psi_q \rangle$ is the probability amplitude of the quasiperiodic wave function at site $\mathbf{r}$, we obtain
\begin{equation}
    [E-W_{diag}(E)] \psi_{\mathbf{r}} = V_{\mathbf{r}}\psi_{\mathbf{r}} + W_{off}(E) \sum_{\mu=1}^n (\psi_{\mathbf{r}+\hat{e}_\mu} + \psi_{\mathbf{r}-\hat{e}_\mu}).
\end{equation}
We multiply both sides of the equation by a rescaling factor $S(E) = t^\kappa / W_{off}(E)$, where $t$ denotes the nearest-neighbor hopping strength between adjacent sites, and the integer $\kappa \ge 2$ represents the shortest hopping path length, defined as the minimum number of nearest-neighbor bonds connecting adjacent quasiperiodic sites. From a perturbation theory perspective, $t^\kappa$ naturally captures the characteristic energy scale of a $\kappa$-th order tunneling process. Furthermore, since $W_{off}(E)$ is exclusively determined by the local geometric structure of a single link, $S(E)$ remains invariant with respect to the global spatial dimension. The resulting equation is: $S(E)[E - W_{diag}(E)] \psi_{\mathbf{r}} = t^\kappa \sum_{\mu=1}^n (\psi_{\mathbf{r}+\hat{e}_\mu} + \psi_{\mathbf{r}-\hat{e}_\mu}) + S(E) V_{\mathbf{r}} \psi_{\mathbf{r}}$.

We define the potential renormalization factor $g(E)=S(E)$ and the effective energy characteristic function $f(E)=g(E)[E - W_{diag}(E)]$. Consequently, we obtain the $n$-dimensional generalized AA equation \cite{SergeAubry1980,Hu2025}
\begin{equation}
  f(E) \psi_{\mathbf{r}} = \sum_{\mu=1}^n[ t^\kappa(\psi_{\mathbf{r} + \hat{e}_\mu} + \psi_{\mathbf{r} - \hat{e}_\mu}) +\\  g(E)V  \cos(2\pi \alpha_\mu {r}_\mu) \psi_{\mathbf{r}}].
    \label{Eq5}
\end{equation}
To reveal the hidden self-duality, we perform a Fourier transformation to momentum space $\psi_{\mathbf{r}} \sim \sum_{k} \text{exp}(i \mathbf{k} \cdot \mathbf{r}) \tilde{\psi}_{\mathbf{k}}$, where $\mathbf{k} = (k_1, k_2, \dots, k_n)$ is the momentum vector and $\tilde{\psi}_\mathbf{k}$ is the corresponding wave function amplitude in momentum space \cite{SergeAubry1980,Wang2021,Goncalves2022}. Substituting this transformation into Eq. (\ref{Eq5}) yields the dual equation $f(E) \tilde{\psi}_{\mathbf{k}} = \sum_{\mu=1}^n \left[ 2 t^\kappa\cos(k_\mu) \tilde{\psi}_{\mathbf{k}} + \frac{g(E)V}{2}  ( \tilde{\psi}_{\mathbf{k} - 2\pi \alpha_\mu \hat{e}_\mu} +  \tilde{\psi}_{\mathbf{k} + 2\pi \alpha_\mu \hat{e}_\mu}) \right]$. By comparing the real-space Eq. (\ref{Eq5}) with its momentum-space counterpart, we obtain the exact mobility edges for $n$-dimensional quasiperiodic networks, as determined by the self-duality condition: 
\begin{equation}
    g(E)V = \pm 2t^\kappa.
    \label{Eq6}
\end{equation}
For this specific class of network models, the exact self-duality condition in Eq. (\ref{Eq6}) provides a well-defined analytical boundary for the mobility edges. It indicates that the critical condition for the localization transition is governed by the interplay between the potential renormalization factor $g(E)$ and the effective hopping strength $t^\kappa$.

 \begin{figure*}[t]
    \centering
    \includegraphics[width=\textwidth]{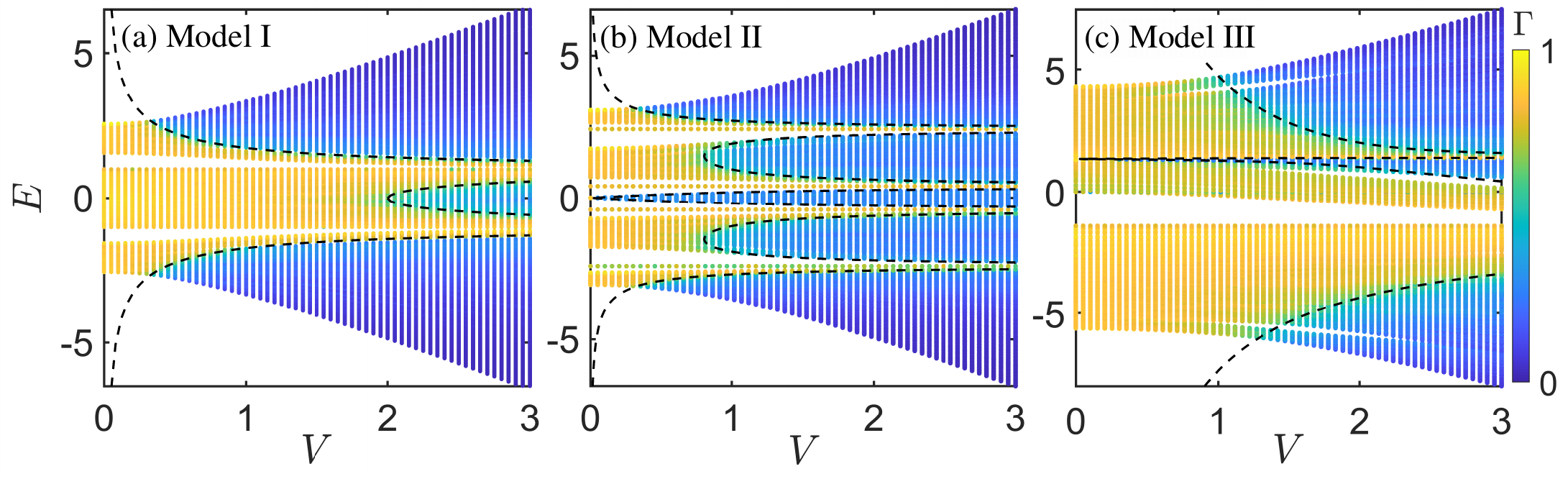}
    \caption{Numerical phase diagrams of energy $E$ versus potential strength $V$ for the three representative models listed in Table~\ref{Table I}: (a) model I (extended Lieb lattice), (b) model II (rhombic-decorated lattice), and (c) model III (densely-connected lattice). The background color indicates the fractal dimension $\Gamma$, while the black dashed lines represent the exact phase boundaries derived from our analytical theory. For panels (a) and (c), the periodic potential parameters are set to $U_k=0$, while for panel (b), the internal potential parameters are set to $U_1=U_2=1$. In all subfigures, the hopping amplitude is $t=1$, the system size of quasiperiodic sites is fixed at $L_x=L_y=34$, and the irrational numbers are set as $\alpha_1=\frac{\sqrt{5}-1}{2}$ and $\alpha_2=\frac{\sqrt{2}}{2}$.}
     \label{fig.2}
\end{figure*}
\section{Geometrically Robust Hidden Self-Duality in 2D}

To validate the universality and precision of the proposed analytical framework under complex connectivity, we construct and systematically investigate three representative classes of 2D quasiperiodic network models in Table~\ref{Table I}, based on the generic structure illustrated in Fig.~\ref{fig.1}(a). 

The models listed in Table~\ref{Table I} exhibit distinct geometric morphologies and localized connectivity patterns in their periodic networks. Model I forms a vertex-decorated extended Lieb lattice, where linear chains bridge quasiperiodic sites via nearest-neighbor hopping \cite{Jiang2019,Bhattacharya2019,Wang2023}. Model II, a rhombic-decorated lattice, incorporates rhombic substructures that introduce additional geometric complexity. Model III, a densely-connected lattice, further increases the geometric connectivity between the periodic networks and the quasiperiodic sites, resulting in a 2D quasiperiodic network with high coordination numbers \cite{Hu2025}. Here, $U_k$ represents the bare onsite energy of the $k$-th site within a single periodic network, distinct from the effective self-energy generated after integration. Indeed, $U_k$ enters the periodic network Hamiltonian $H_p$ and consequently modifies $H_c(E-H_p)^{-1}H_c^\dagger$, thereby changing the functions $g(E)$ and $f(E)$ and hence the positions of the mobility edges. 

Despite significant differences in geometric morphology, coordination number, and coupling mechanisms, rigorously integrating out the periodic site wave functions $\psi_p$ allows the original multicomponent Schrödinger equation for all three models to be exactly mapped onto a unified 2D generalized AA equation within the quasiperiodic site subspace: 
\begin{equation}
    \begin{aligned}
        g(E)V_{i,j}\psi_{i,j}+t^\kappa(\psi_{i+1,j}+\psi_{i-1,j}+\psi_{i,j-1}+\psi_{i,j+1})\\=f(E)\psi_{i,j},
    \end{aligned}
\end{equation}
\begin{table*}[t]
    \renewcommand{\arraystretch}{2.2}
    \centering
    \caption{Dimensional evolution of the potential renormalization factor $g(E)$ and the effective energy characteristic function $f_{nD}(E)$ across the three representative quasiperiodic network models. The construction of the $g(E)$ cells highlights its strict dimension-independence, whereas $f_{nD}(E)$ scales explicitly with the spatial dimension $n$.}
    
    \resizebox{\textwidth}{!}{ 
    \begin{tabular}{|c|c|c|c|c|c|c|}
        \hline
        \multirow{2}{*}{\textbf{Dim.}} & \multicolumn{2}{c|}{\textbf{Model I}} & \multicolumn{2}{c|}{\textbf{Model II}} & \multicolumn{2}{c|}{\textbf{Model III}} \\
        \cline{2-7}
        & $g(E)$ & $f(E)$ & $g(E)$ & $f(E)$ & $g(E)$ & $f(E)$ \\
        \hline
        \textbf{1D} & \multirow{4}{*}{$\displaystyle t^2 - E^2$} & $\displaystyle E(3t^2 - E^2)$ & \multirow{4}{*}{$\displaystyle \frac{E^4 - E^2(U_1^2 + U_2^2 + 4t^2) + U_1^2 U_2^2}{2E}$} & $\displaystyle \frac{E^4 - E^2(U_1^2 + U_2^2 + 6t^2) + U_1^2 U_2^2 + 2t^2 U_2^2 + 4t^4}{2}$ & \multirow{4}{*}{$\displaystyle \frac{E^2 - 2t^2}{3E - 4t}$} & $\displaystyle \frac{E^3 - 8Et^2 + 8t^3}{3E - 4t}$ \\
        \cline{1-1}\cline{3-3}\cline{5-5}\cline{7-7}
        \textbf{2D} & & $\displaystyle E(5t^2 - E^2)$ & & $\displaystyle \frac{E^4 - E^2(U_1^2 + U_2^2 + 8t^2) + U_1^2 U_2^2 + 4t^2 U_2^2 + 8t^4}{2}$ & & $\displaystyle \frac{E^3 - 14Et^2 + 16t^3}{3E - 4t}$ \\
        \cline{1-1}\cline{3-3}\cline{5-5}\cline{7-7}
        \textbf{3D} & & $\displaystyle E(7t^2 - E^2)$ & & $\displaystyle \frac{E^4 - E^2(U_1^2 + U_2^2 + 10t^2) + U_1^2 U_2^2 + 6t^2 U_2^2 + 12t^4}{2}$ & & $\displaystyle \frac{E^3 - 20Et^2 + 24t^3}{3E - 4t}$ \\
        \cline{1-1}\cline{3-3}\cline{5-5}\cline{7-7}
        \textbf{$n$D} & & $\displaystyle (t^2 - E^2)E - 2n(-Et^2)$ & & $\displaystyle \frac{E^4 - E^2(U_1^2 + U_2^2 + 4t^2) + U_1^2 U_2^2}{2} - n t^2 (E^2 - U_2^2 - 2t^2)$ & & $\displaystyle \frac{E^3 - 2t^2 E - 2nt^2(3E - 4t)}{3E - 4t}$ \\
        \hline
    \end{tabular}
    }
    \label{Table II}
\end{table*}
where $\psi_{i,j}$ denotes the normalized wave function amplitude at quasiperiodic site $(i,j)$, and the quasiperiodic potential is given by $V_{i,j}=V[\cos(2\pi\alpha_1 i)+\cos(2\pi\alpha_2 j)]$. The specific analytical expressions for the effective energy characteristic function $f(E)$ and the potential renormalization factor $g(E)$ for each model are provided in Table~\ref{Table I}. 

The analytical results can be numerically verified by calculating the fractal dimension, which is defined as $\Gamma=-\lim_{N\to\infty}\ln(IPR)/\ln{N}$ \cite{Wang2023}. Here, the inverse participation ratio is given by $IPR=\sum^{N}_{s=1}|\psi_{s}|^4$, where $\psi_{s}$ denotes the normalized wave function amplitude at the $s$-th site of the original full lattice \cite{Evers2008} and $N=L_x\times L_y\times (1+2m)$ represents the total number of sites. Specifically, $L_x$ and $L_y$ denote the number of unit cells along the $x$ and $y$ spatial directions, respectively. In our lattice structure, each unit cell consists of exactly one quasiperiodic primary site and the periodic networks along the positive spatial directions. Here, $m$ represents the number of periodic sites within a single periodic network. The values $\Gamma \to 1$ and $\Gamma \to 0$ correspond to extended and localized states, respectively, while $0 < \Gamma < 1$ indicates multifractal states.

Fig.~\ref{fig.2} displays $\Gamma$ of different eigenstates as a function of $V$ and the corresponding eigenvalues $E$ for distinct 2D models listed in Table~\ref{Table I}. A detailed finite-size scaling analysis is provided in Appendix \ref{Appendix B:}. The results show that the states in the yellow and blue regions are extended and localized, respectively. They are separated by the black dashed lines, which represent the mobility edges described by Eq. (\ref{Eq6}). In Fig.~\ref{fig.2}, setting $U_k=0$ for Models I and III provides simple symmetric cases, while $U_1=U_2=t$ (with $t=1$ in our calculations) for Model II illustrates a finite-onsite-energy configuration. This confirms the exact mapping relies solely on the structural conditions in Sec. II rather than specific parameters. Notably, at $E=0$ in Fig.~\ref{fig.2}(b) and $E=4t/3$ in Fig.~\ref{fig.2}(c), the system remains extended only at $V=0$ and undergoes a sharp localization transition for any $V>0$. As shown in Table~\ref{Table I}, these specific energy points coincide with the poles of $g(E)$ (where $W_{off}(E)=0$). At these singularities, destructive interference completely halts inter-site tunneling, while the diverging effective potential $g(E)V$ infinitely amplifies any non-zero quasiperiodic perturbation, ultimately driving the system into deep localization.

These results demonstrate that variations in the connectivity structure of 2D quasiperiodic networks do not compromise the solvability of the system or alter the fundamental criterion for mobility edges. Instead, the specific geometric details are encoded into the effective hopping strength $t^\kappa$ and the functional forms of $g(E)$ and $f(E)$ through the integration-out process. 

\section{Dimensional Evolution of Effective Characteristics}

We now proceed to investigate the universality of this theoretical framework in higher spatial dimensions by considering the three-dimensional network schematically shown in Fig.~\ref{fig.1}(b). We find that, for the specific subset of models considered herein, the dimensional evolution of the potential renormalization factor $g(E)$ and the effective energy characteristic function $f(E)$ is strictly governed by the independence of multi-path scattering processes and the coordination-number-dependent scaling laws, respectively.

Given $g(E) = t^\kappa / W_{off}(E)$, the strict decoupling along different spatial directions of periodic links ensures that $W_{off}(E)$ depends solely on the periodic network of a single bond. Since both this local geometry and the associated hopping path $\kappa$ remain invariant across dimensions, $g(E)$ is strictly dimension-independent, thereby preserving the exact self-duality criterion $g(E)V = \pm 2t^\kappa$ universally.
\begin{figure}[!t] 
    \centering
    \includegraphics[width=\columnwidth]{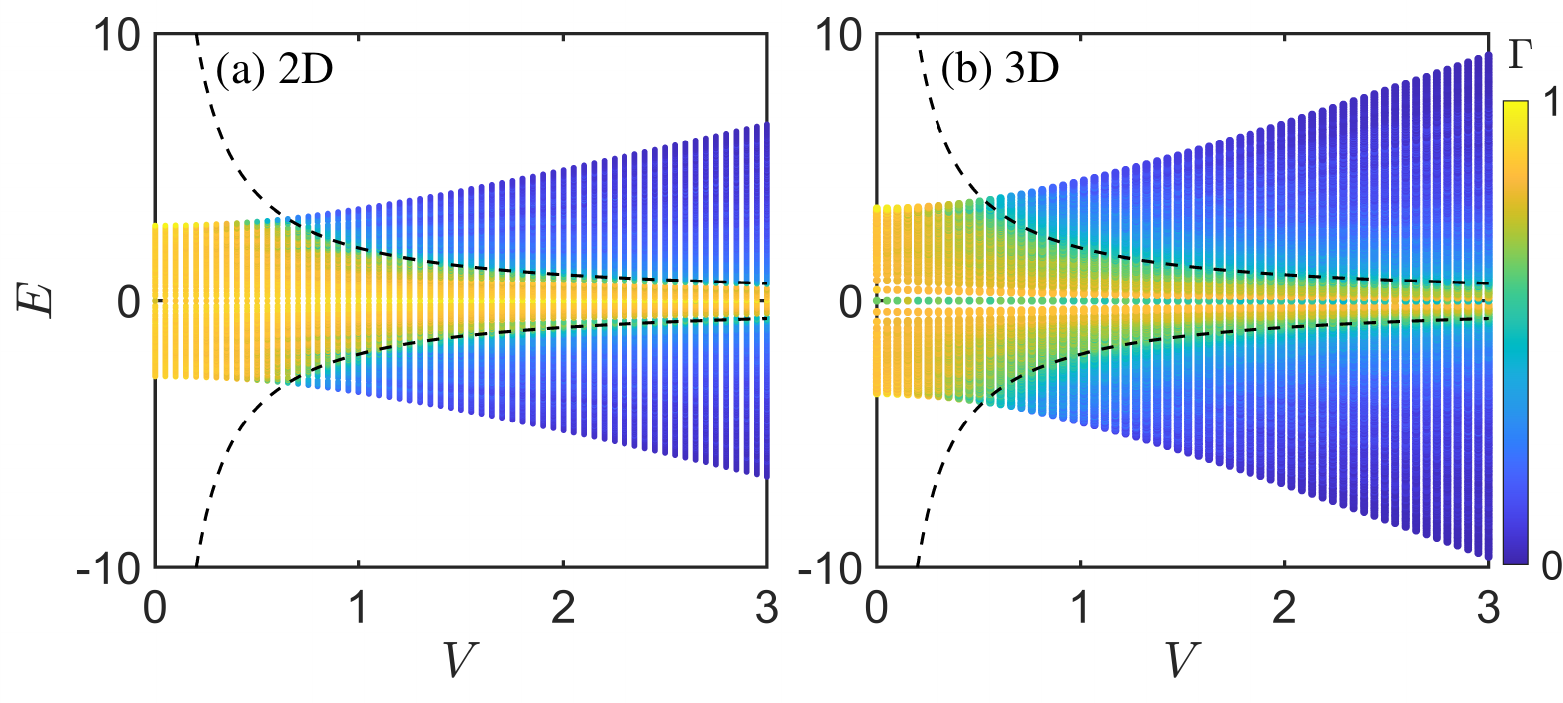}
    \caption{Phase diagrams of energy $E$ versus potential strength $V$ for Lieb lattices. (a) The 2D Lieb lattice with a system size of $L_x=L_y=34$. (b) The 3D Lieb lattice with a system size of $L_x=L_y=L_z=13$. The parameters for both models are set to $\alpha_1=\frac{\sqrt{5}-1}{2}$, $\alpha_2=\frac{\sqrt{2}}{2}$, $\alpha_3=\frac{\pi}{4}$, and $t=1$.}
     \label{fig.3}
\end{figure}
In sharp contrast, $f(E)$ exhibits an explicit dimensional dependence originating from the linear accumulation of local self-energies across the multi-dimensional network. For an $n$-dimensional hypercubic lattice with a coordination number $Z_n = 2n$, the mutually independent periodic branches ensure that the total diagonal self-energy satisfies $W_{diag}(E) = Z_n \cdot w_{unit}(E)$, where $w_{unit}(E)$ is the diagonal self-energy contributed by a single periodic link. Substituting this relation into $f(E) = g(E)[E - W_{diag}(E)]$, we directly obtain $f(E) = g(E)E - Z_n \cdot [g(E) w_{unit}(E)]$. Defining the renormalized single-link self-energy as $\tilde{\Sigma}_{unit}(E)=g(E) w_{unit}(E)$ yields the universal scaling formula extending $f(E)$ from 1D to arbitrary $n$ dimensions:
\begin{equation}
    f_{nD}(E) = g(E)E - Z_n \cdot \tilde{\Sigma}_{unit}(E). 
    \label{Eq8}
\end{equation}

Physically, this formula demonstrates that the effective energy characteristic in multi-dimensional systems does not arise from higher-order nonlinearities but rather strictly obeys a linear superposition principle, where the rescaled self-energy $\tilde{\Sigma}_{unit}(E)$ from individual independent links accumulates additively, governed solely by the coordination number $Z_n$.

We systematically derive the key characteristic functions, $g(E)$ and $f(E)$, for the three models introduced in Table~\ref{Table I} across one-, two-, and three-dimensional systems, as summarized in Table~\ref{Table II}. All results can be cast into the universal scaling form of Eq.~(\ref{Eq8}), demonstrating that, despite their complex internal structures, the mobility-edge criterion remains strictly dimension-independent, whereas the effective energy characteristic evolves systematically with spatial dimension.

To visually verify the dimension-independence of $g(E)$, we select the standard Lieb lattice \cite{Lieb1989,Liu2020,Devakul2017} as the $n$-dimensional baseline model. For this model, $g(E)$ remains invariant across dimensions, yielding $g(E)_{1D} = g(E)_{2D} = g(E)_{3D} = E$. Conversely, $f(E)$ depends on the spatial dimension $n$, taking the forms $f_{1D}(E) = E^2 - 2t^2$, $f_{2D}(E) = E^2 - 4t^2$, and $f_{3D}(E) = E^2 - 6t^2$. Fig.~\ref{fig.3}(a) and~\ref{fig.3}(b) present the numerical phase diagrams for the 2D and 3D Lieb lattices, respectively. Despite the reconfiguration of the energy spectrum driven by $f(E)$, the mobility edges in both 2D and 3D cases are consistent with the identical self-duality criterion $g(E)V = \pm2t^\kappa$. 

Notably, at $E=0$, the Lieb lattice features macroscopic degenerate flat bands comprised of compact localized states (CLS) \cite{Leykam2018}. Since these CLS reside entirely within the periodic networks with $\psi_q = 0$, they are immune to the quasiperiodic potential as proven in Appendix \ref{Appendix B:}. Consequently, the apparent critical behaviors observed at $E=0$ in Fig. \ref{fig.3}(b) are not driven by the Anderson localization mechanism, but stem from exact diagonalization solvers generating arbitrary orthogonal linear combinations within this highly degenerate subspace. Due to the restricted system size ($L=13$), these widespread superpositions exhibit multifractal-like behavior, while in fact remaining localized.

Our analysis reveals that, within our proposed analytical framework, mobility edges in this specific class of high-dimensional quasiperiodic networks can be systematically understood through an exact mapping onto effective high-dimensional AA models, where dimensional effects manifest via the linear superposition of local self-energies. Consequently, localization transitions in this targeted subset of high-dimensional quasiperiodic systems across arbitrary dimensions can be uniformly characterized, providing a clear theoretical foundation for understanding ME mechanisms in these complex lattices.

\section{Conclusion}
In summary, we generalize the 1D hidden self-duality framework \cite{Hu2025} to a class of higher-dimensional quasiperiodic network models. When the periodic network serves exclusively as bond-decorating links between quasiperiodic sites and the quasiperiodic potential is spatially separable, the original system can be mapped exactly onto an effective higher-dimensional generalized AA model defined on the quasiperiodic subspace by integrating out the periodic degrees of freedom. This mapping restores the hidden self-duality of the effective Hamiltonian and yields analytical criteria for the exact mobility edges. A notable consequence is that the potential renormalization factor $g(E)$ remains invariant with spatial dimension, whereas the effective energy function $f(E)$ is determined by the lattice geometry through the spatial dimension and coordination number. These results reveal that the localization transition in this class of high-dimensional quasiperiodic networks can be characterized within a unified effective AA framework. Analytical predictions for representative two- and three-dimensional models agree quantitatively with numerical phase diagrams.

This framework provides a systematic method to engineer and predict localization features of complex high-dimensional lattices. Besides, since eliminating redundant degrees of freedom has recently been shown to enable the construction of multifractality-enriched mobility edges \cite{Li2026}, the hidden self-duality mechanism may also be extended to obtain exact multifractality-enriched mobility edges for such systems in higher dimensions. Furthermore, incorporating synthetic gauge fields into these networks provides an analytically tractable foundation for exploring the interplay between flat-band topology and localization \cite{Bhattacharya2019}. Relatedly, the mathematical procedure of integrating out local degrees of freedom to reveal latent symmetries offers a structural pathway to engineer topological phases \cite{Mei2020,Zhang2024}, such as topological Anderson insulators \cite{Lin2026}. Additionally, applying Floquet periodic driving to these models facilitates the dynamic control of mobility edges \cite{Gao2025a,Zhou2021} and higher-order topological states \cite{Zhang2025,Qian2025} in non-equilibrium regimes.

\acknowledgments
This work was supported by the National Key Research and Development Program of China (Grant No. 2022YFA1405300), the Open Fund of Key Laboratory of Atomic and Subatomic Structure and Quantum Control (Ministry of Education) and the Guangdong Provincial Quantum Science Strategic Initiative (Grant No. GDZX2204003). Jia-Hui Zhang and Shan-Zhong Li contributed equally to this work. 

\appendix
\section{Derivations of the effective Hamiltonians for 2D quasiperiodic networks}
\label{Appendix A}

In this appendix, we provide the detailed derivations for the three 2D quasiperiodic network models summarized in Table~\ref{Table I}.

\subsection{Model I: Extended Lieb Lattice}

In the 2D Extended Lieb lattice, adjacent primary sites are bridged by two intermediate periodic sites with $U_k = 0$ and nearest-neighbor hopping amplitude $t$. For the $x$-directional bond between primary sites $\psi_{i,j}$ and $\psi_{i+1,j}$, the intermediate sites $d_{1,x}$ and $d_{2,x}$ yield coupled Schrödinger equations in matrix form
\begin{equation}
    \begin{pmatrix} E & t \\ t & E \end{pmatrix} 
    \begin{pmatrix} d_{1,x} \\ d_{2,x} \end{pmatrix} 
    = -t \begin{pmatrix} \psi_{i,j} \\ \psi_{i+1,j} \end{pmatrix}.
\end{equation}
Consequently, we obtain the solutions for both intermediate sites
\begin{equation}
    \begin{aligned}
        d_{1,x} &= -\frac{t}{E^2 - t^2} (E \psi_{i,j} - t \psi_{i+1,j}), \\
        d_{2,x} &= -\frac{t}{E^2 - t^2} (E \psi_{i+1,j} - t \psi_{i,j}).
    \end{aligned}
\end{equation}
The stationary equation for site $\psi_{i,j}$ integrates hopping feedback from its four nearest neighbors, which reads
\begin{equation}
    E\psi_{i,j} = V_{i,j}\psi_{i,j} - t (d_{1,x} + d_{2,x-1} + d_{1,y} + d_{2,y-1}).
    \label{A3}
\end{equation}
Shifting the spatial index $x\to x-1$ in the analytical solution yields $d_{2,x-1}=-\frac{t}{E^2-t^2}(E\psi_{i,j}-t\psi_{i-1,j})$, similarly for the $y$-direction. Substituting the resulting equations into Eq. ~(\ref{A3}), we can obtain
\begin{equation}
    (E-\frac{4t^2 E}{E^2 - t^2})\psi_{i,j} = V_{i,j}\psi_{i,j}  - \frac{t^3}{E^2 - t^2} \sum_{\langle i',j' \rangle} \psi_{i',j'},
\end{equation}
where $\sum_{\langle i',j' \rangle}$ denotes the sum over the four nearest-neighbor primary sites. With shortest hopping path length $\kappa=3$, we multiply by the rescaling factor $S(E) = t^2 - E^2$, resulting in
\begin{equation}
    (5t^2 E - E^3) \psi_{i,j} = (t^2 - E^2) V_{i,j} \psi_{i,j} + t^3 \sum_{\langle i',j' \rangle} \psi_{i',j'}.
\end{equation}
By matching this with the generalized form $f(E)\psi_{i,j} = g(E)V_{i,j}\psi_{i,j} + t^\kappa \sum \psi_{i',j'}$, we extract:
\begin{equation}
     \begin{aligned}
    g(E) &= t^2 - E^2, \\
    f(E) &= E(5t^2 - E^2).
    \end{aligned}
\end{equation}

\subsection{Model II: Rhombic-decorated Lattice}

In Model II, the periodic network is a four-site rhombus (potentials $U_1, U_2, -U_1, -U_2$) labeled $d_{1,x}$ to $d_{4,x}$. The inner nodes $d_{2,x}$ and $d_{4,x}$ can be algebraically eliminated, yielding $d_{2,x} + d_{4,x} = \frac{-2tE}{E^2 - U_2^2} (d_{1,x} + d_{3,x})$. Defining the internal coupling constant $K(E)=\frac{2t^2 E}{E^2 - U_2^2}$, the equations for the outer nodes $d_{1,x}$ and $d_{3,x}$ are reduced to a matrix system
\begin{equation}
    \begin{pmatrix} E - U_1 - K & -K \\ -K & E + U_1 - K \end{pmatrix}
    \begin{pmatrix} d_{1,x} \\ d_{3,x} \end{pmatrix}
    = -t \begin{pmatrix} \psi_{i,j} \\ \psi_{i+1,j} \end{pmatrix}.
\end{equation}
The determinant of this matrix is $D = \frac{Num_D}{E^2 - U_2^2}$, where the numerator is $Num_D = E^4 - E^2(U_1^2 + U_2^2 + 4t^2) + U_1^2 U_2^2$. Matrix inversion yields the explicit solutions:
\begin{equation}
    \begin{aligned}
        d_{1,x} &= -\frac{t}{D} \left[ (E + U_1 - K) \psi_{i,j} + K \psi_{i+1,j} \right], \\
        d_{3,x-1} &= -\frac{t}{D} \left[ (E - U_1 - K) \psi_{i,j} + K \psi_{i-1,j} \right].
    \end{aligned}
\end{equation}
The primary equation is $E \psi_{i,j} = V_{i,j} \psi_{i,j} - t \sum_{\nu \in \{x,y\}} (d_{1,\nu} + d_{3,\nu-1})$. Substituting the solutions and summing over the four spatial directions, we obtain
\begin{equation}
    ( E - \frac{4t^2(E - K)}{D}) \psi_{i,j} = V_{i,j} \psi_{i,j} + \frac{2t^4 E}{Num_D} \sum_{\langle i',j' \rangle} \psi_{i',j'}.
\end{equation}
With shortest hopping path length $\kappa=4$, we multiply both sides by $S(E) = \frac{Num_D}{2E}$, resulting in
\begin{equation}
    \begin{aligned}
        \frac{Num_D}{2E} \left( E - \frac{4t^2(E - K)}{D} \right) \psi_{i,j} = \frac{Num_D}{2E} V_{i,j} \psi_{i,j} \\+ t^4 \sum_{\langle i',j' \rangle} \psi_{i',j'}.
    \end{aligned}
\end{equation}
Extracting the coefficients, we obtain $g(E)$ and the explicitly expanded $f(E)$:
\begin{equation}
    \begin{aligned}
        g(E) &= \frac{E^4 - E^2(U_1^2 + U_2^2 + 4t^2) + U_1^2 U_2^2}{2E}, \\
        f(E) &= \frac{E^4 - E^2(U_1^2 + U_2^2 + 8t^2) + U_1^2 U_2^2 + 4U_2^2t^2 + 8t^4 }{2}.
    \end{aligned}
\end{equation}

\subsection{Model III: Densely-connected Lattice}

In Model III, bonds between $\psi_{i,j}$ and $\psi_{i+1,j}$ contain three intermediate sites ($d_{1,x}, d_{2,x}, d_{3,x}$), where $d_{1,x} = d_{3,x}$ due to reflection symmetry. Defining $\tilde{\psi}_x = \psi_{i,j} + \psi_{i+1,j}$, the intermediate sites equations form a matrix
\begin{equation}
    \begin{pmatrix} E & t \\ 2t & E \end{pmatrix} 
    \begin{pmatrix} d_{1,x} \\ d_{2,x} \end{pmatrix} 
    = -t \tilde{\psi}_x \begin{pmatrix} 1 \\ 1 \end{pmatrix}.
\end{equation}
The determinant of this matrix is $D = E^2 - 2t^2$, from which we readily get $d_{1,x}=\frac{t(t-E)}{E^2 - 2t^2}\tilde{\psi}_x$ and $d_{2,x} = \frac{t(2t - E)}{E^2 - 2t^2} \tilde{\psi}_x$. The total feedback generated from this single bond is:
\begin{equation}
    2d_{1,x} + d_{2,x} = \frac{4t^2 - 3tE}{E^2 - 2t^2} (\psi_{i,j} + \psi_{i+1,j}).
\end{equation}
Substituting this total bond contribution into the primary equation $E \psi_{i,j} = V_{i,j} \psi_{i,j} - t \sum_{\langle i',j' \rangle} (2d_{1,\nu} + d_{2,\nu})$, we obtain
\begin{equation}
    \begin{aligned}
        \left( E - \frac{12t^2 E - 16t^3}{E^2 - 2t^2} \right) \psi_{i,j} = V_{i,j} \psi_{i,j} \\+ \frac{3t^2 E - 4t^3}{E^2 - 2t^2} \sum_{\langle i',j' \rangle} \psi_{i',j'}.
    \end{aligned}
\end{equation}
With shortest hopping path length $\kappa=2$, we multiply by the rescaling factor $S(E) = \frac{E^2 - 2t^2}{3E - 4t}$, resulting in
\begin{equation}
    \begin{aligned}
        \frac{E^3 - 14Et^2 + 16t^3}{3E - 4t} \psi_{i,j} = \frac{E^2 - 2t^2}{3E - 4t} V_{i,j} \psi_{i,j} \\+ t^2 \sum_{\langle i',j' \rangle} \psi_{i',j'}.
    \end{aligned}
\end{equation}
Consequently, we obtain $g(E)$ and expanded $f (E)$:
\begin{equation}
    \begin{aligned}
        g(E) &= \frac{E^2 - 2t^2}{3E - 4t}, \\
        f(E) &= \frac{E^3 - 14Et^2 + 16t^3}{3E - 4t}.
    \end{aligned}
\end{equation}

\begin{figure*}[t]
    \centering
    \includegraphics[width=0.95\textwidth]{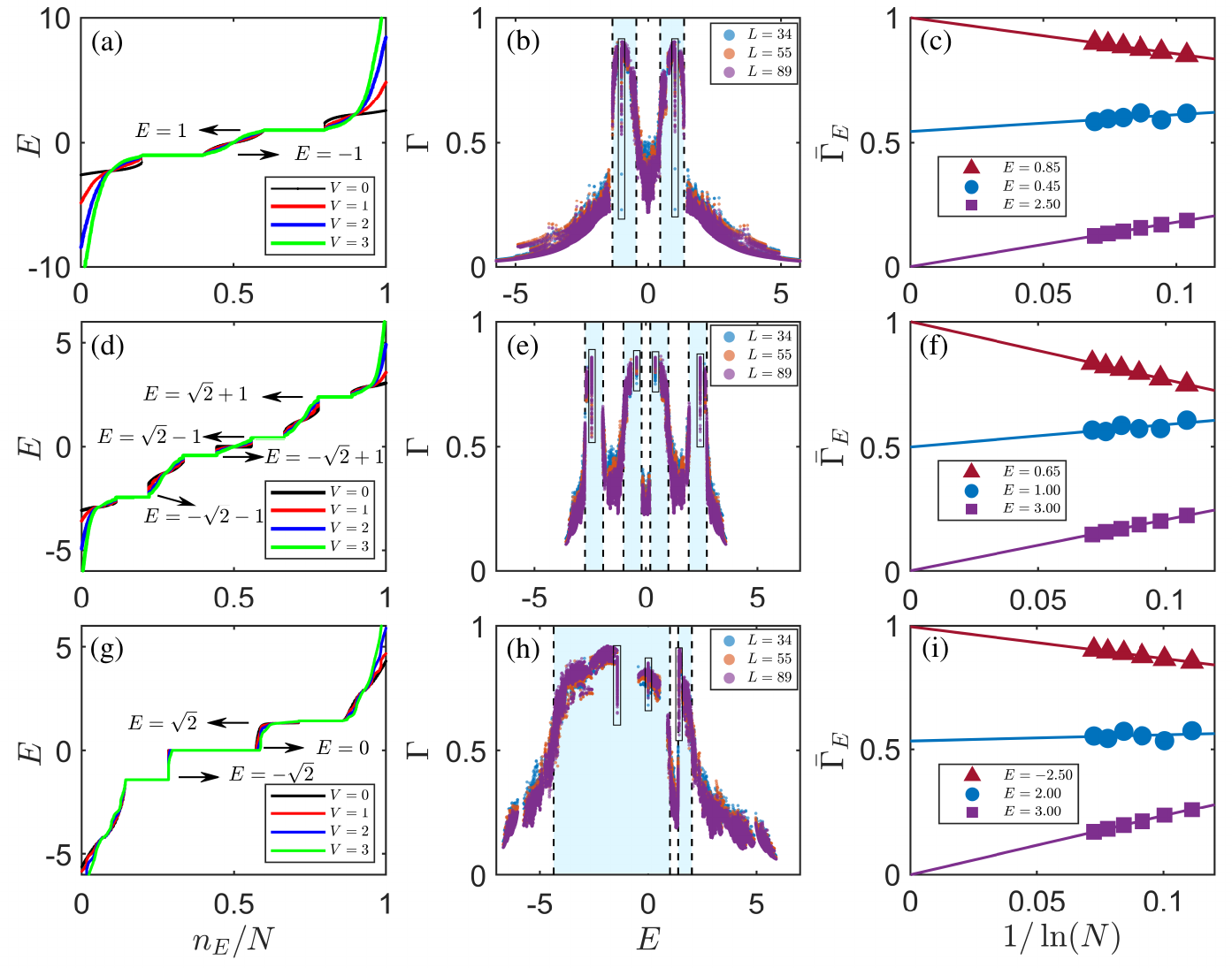}
    \caption{(a, d, g) Eigenvalues as a function of the normalized state index $n_E/N$ for Model I, II, and III in Table~\ref{Table I}, respectively. (b, e, h) The fractal dimension $\Gamma$ versus eigenenergy $E$ across different system sizes, evaluated at fixed quasiperiodic potential amplitude of $V=2.5,1,2$ for Model I, II, III, respectively. (c, f, i) The averaged fractal dimension $\bar{\Gamma}_E$ versus $1/\ln(N)$ for representative extended (triangles), critical (circles), and localized (squares) states extracted from (b), (e), and (h) at the indicated energies for Models I, II, and III respectively, with solid lines indicating extrapolations to the thermodynamic limit. In all panels, $t=1$, and all other parameters are identical to those in Fig. \ref{fig.2}.}
    \label{fig.4}
\end{figure*}

\section{Flat bands and finite-size scaling in the 2D models}
\label{Appendix B:}
In this section, we conduct a finite-size analysis on the three two-dimensional models presented in Table ~\ref{Table I} to rigorously verify the accuracy of the mobility edges in the thermodynamic limit and elucidate the local impact of flat bands on the phase diagram.

The complete Schrödinger equation of the system is given by:
\begin{equation}
    \begin{pmatrix} H_q & H_c \\ H_c^\dagger & H_p \end{pmatrix} \begin{pmatrix} \psi_q \\ \psi_p \end{pmatrix} = E \begin{pmatrix} \psi_q \\ \psi_p \end{pmatrix}.
    \label{B1}
\end{equation}
If flat bands are completely immune to the quasiperiodic potential $V_\mathbf{r}$, their wave function amplitudes on the quasiperiodic primary sites must strictly vanish ($\psi_q = 0$). Substituting this condition into Eq. (\ref{B1}) gives two independent equations:
\begin{equation}
\begin{aligned}
    H_p \psi_p &= E \psi_p, \\
    H_c \psi_p &= 0.
\end{aligned}
\end{equation}
These two equations define the physical nature of the flat bands in our proposed models, which is fundamentally rooted in the local geometric connectivity of the lattice \cite{Sutherland1986}. The former indicates that the flat band energy $E$ must be an eigenvalue of the periodic network block matrix $H_p$, whose dimension is equal to the total number of periodic sites $m$ in a single network, resulting in $m$ dispersionless flat bands. The latter dictates that the wave functions on the periodic network undergo destructive interference when hopping to neighboring primary sites, forming macroscopic degenerate CLS entirely trapped within the periodic bonds \cite{Leykam2018}. As verified in Fig. \ref{fig.4}(a, d, g): Model I ($m=2$) generates 2 flat bands at $E = \pm t$, Model II ($m=4$) corresponds to 4 flat bands at $\pm\sqrt{2}+1, \pm\sqrt{2}-1$, and Model III ($m=3$) produces 3 flat bands at $E = 0, \pm\sqrt{2}t$. 

The impact of these flat bands on the localization transition is visually evident yet physically distinct. In Fig. \ref{fig.4}(b, e, h), while theoretical mobility edges separate extended (blue) and localized (white) phases, $\Gamma$ exhibits pronounced dips and fluctuations at the flat-band energies highlighted by black boxes. This anomaly is not a physical breakdown of the extended phase, but a direct consequence of macroscopic degeneracy. Because these highly degenerate CLS are immune to the quasiperiodic potential, numerical solvers output arbitrary linear combinations that artificially inflate their spatial distribution. Consequently, the resulting $\Gamma$ dips reflect the basis-selection ambiguity of the degenerate subspace, which are fundamentally independent of the quasiperiodic-driven localization transition.

Beyond the flat-band anomalies, the $\Gamma$ faithfully captures the localization transitions in Fig. \ref{fig.4}(b, e, h), remaining high in the blue extended regions and approaching zero in the white localized regions. However, since state scaling within computationally accessible limits ($L=34$ to $89$) exhibits a clear downward trend with increasing size in the localized regime while remaining less pronounced in the extended region, lacking sufficient prominence to confirm macroscopic convergence, we calculate the averaged fractal dimension $\overline{\Gamma}_E$ using 150 targeted eigenstates. In Fig. \ref{fig.4}(c, f, i), as $N \to \infty$, $\overline{\Gamma}_E$ converges to 1 for extended states and 0 for localized states, while critical states at the mobility edges exhibit multifractal characteristics by stably converging to fractional values near 0.5. Ultimately, by explicitly isolating numerical artifacts and demonstrating macroscopic convergence, this analysis conclusively verifies the exactness of our theoretically predicted high-dimensional mobility edges in the thermodynamic limit. 

\bibliography{main}

@article{Abrahams1979,
  title = {Scaling Theory of Localization: {{Absence}} of Quantum Diffusion in Two Dimensions},
  shorttitle = {Scaling Theory of Localization},
  author = {Abrahams, E. and Anderson, P. W. and Licciardello, D. C. and Ramakrishnan, T. V.},
  year = 1979,
  month = mar,
  journal = {Phys. Rev. Lett.},
  volume = {42},
  number = {10},
  pages = {673--676},
  publisher = {American Physical Society},
  doi = {10.1103/PhysRevLett.42.673},
  urldate = {2026-03-01}
}

@article{Anderson1958,
  title = {Absence of Diffusion in Certain Random Lattices},
  author = {Anderson, P. W.},
  year = 1958,
  month = mar,
  journal = {Phys. Rev.},
  volume = {109},
  number = {5},
  pages = {1492--1505},
  publisher = {American Physical Society},
  doi = {10.1103/PhysRev.109.1492},
  urldate = {2026-03-01}
}

@article{Avila2015,
  title = {Global Theory of One-Frequency {{Schr\"odinger}} Operators},
  author = {Avila, Artur},
  year = 2015,
  month = jan,
  journal = {Acta Math.},
  volume = {215},
  number = {1},
  pages = {1--54},
  publisher = {Institut Mittag-Leffler},
  issn = {0001-5962, 1871-2509},
  doi = {10.1007/s11511-015-0128-7},
  urldate = {2026-03-02}
}

@article{Bhattacharya2019,
  title = {Flat Bands and Nontrivial Topological Properties in an Extended {{Lieb}} Lattice},
  author = {Bhattacharya, Ankita and Pal, Biplab},
  year = 2019,
  month = dec,
  journal = {Phys. Rev. B},
  volume = {100},
  number = {23},
  pages = {235145},
  publisher = {American Physical Society},
  doi = {10.1103/PhysRevB.100.235145},
  urldate = {2026-03-06}
}

@article{Biddle2009,
  title = {Localization in One-Dimensional Incommensurate Lattices beyond the {{Aubry-Andr\'e}} Model},
  author = {Biddle, J. and Wang, B. and Priour, D. J. and Das Sarma, S.},
  year = 2009,
  month = aug,
  journal = {Phys. Rev. A},
  volume = {80},
  number = {2},
  pages = {021603},
  publisher = {American Physical Society},
  doi = {10.1103/PhysRevA.80.021603},
  urldate = {2026-03-02}
}

@article{Biddle2010,
  title = {Predicted Mobility Edges in One-Dimensional Incommensurate Optical Lattices: {{An}} Exactly Solvable Model of {{Anderson}} Localization},
  shorttitle = {Predicted Mobility Edges in One-Dimensional Incommensurate Optical Lattices},
  author = {Biddle, J. and Das Sarma, S.},
  year = 2010,
  month = feb,
  journal = {Phys. Rev. Lett.},
  volume = {104},
  number = {7},
  pages = {070601},
  publisher = {American Physical Society},
  doi = {10.1103/PhysRevLett.104.070601},
  urldate = {2026-03-02}
}

@article{Billy2008,
  title = {Direct Observation of {{Anderson}} Localization of Matter Waves in a Controlled Disorder},
  author = {Billy, Juliette and Josse, Vincent and Zuo, Zhanchun and Bernard, Alain and Hambrecht, Ben and Lugan, Pierre and Cl{\'e}ment, David and {Sanchez-Palencia}, Laurent and Bouyer, Philippe and Aspect, Alain},
  year = 2008,
  month = jun,
  journal = {Nature},
  volume = {453},
  number = {7197},
  pages = {891--894},
  publisher = {Nature Publishing Group},
  issn = {1476-4687},
  doi = {10.1038/nature07000},
  urldate = {2026-02-22}
}

@article{Bordia2017,
  title = {Probing Slow Relaxation and Many-Body Localization in Two-Dimensional Quasiperiodic Systems},
  author = {Bordia, Pranjal and L{\"u}schen, Henrik and Scherg, Sebastian and Gopalakrishnan, Sarang and Knap, Michael and Schneider, Ulrich and Bloch, Immanuel},
  year = 2017,
  month = nov,
  journal = {Phys. Rev. X},
  volume = {7},
  number = {4},
  pages = {041047},
  publisher = {American Physical Society},
  doi = {10.1103/PhysRevX.7.041047},
  urldate = {2026-03-01}
}

@article{Danieli2015,
  title = {Flat-Band Engineering of Mobility Edges},
  author = {Danieli, Carlo and Bodyfelt, Joshua D. and Flach, Sergej},
  year = 2015,
  month = jun,
  journal = {Phys. Rev. B},
  volume = {91},
  number = {23},
  pages = {235134},
  publisher = {American Physical Society},
  doi = {10.1103/PhysRevB.91.235134},
  urldate = {2026-03-05}
}

@article{DasSarma1988,
  title = {Mobility Edge in a Model One-Dimensional Potential},
  author = {Das Sarma, S. and He, Song and Xie, X. C.},
  year = 1988,
  month = oct,
  journal = {Phys. Rev. Lett.},
  volume = {61},
  number = {18},
  pages = {2144--2147},
  publisher = {American Physical Society},
  doi = {10.1103/PhysRevLett.61.2144},
  urldate = {2026-03-03}
}

@article{Deng2019,
  title = {One-Dimensional Quasicrystals with Power-Law Hopping},
  author = {Deng, X. and Ray, S. and Sinha, S. and Shlyapnikov, G. V. and Santos, L.},
  year = 2019,
  month = jul,
  journal = {Phys. Rev. Lett.},
  volume = {123},
  number = {2},
  pages = {025301},
  publisher = {American Physical Society},
  doi = {10.1103/PhysRevLett.123.025301},
  urldate = {2026-03-05}
}

@article{Devakul2017,
  title = {Anderson Localization Transitions with and without Random Potentials},
  author = {Devakul, Trithep and Huse, David A.},
  year = 2017,
  month = dec,
  journal = {Phys. Rev. B},
  volume = {96},
  number = {21},
  pages = {214201},
  publisher = {American Physical Society},
  doi = {10.1103/PhysRevB.96.214201},
  urldate = {2026-03-02}
}

@article{Duncan2024,
  title = {Critical States and Anomalous Mobility Edges in Two-Dimensional Diagonal Quasicrystals},
  author = {Duncan, Callum W.},
  year = 2024,
  month = jan,
  journal = {Phys. Rev. B},
  volume = {109},
  number = {1},
  pages = {014210},
  publisher = {American Physical Society},
  doi = {10.1103/PhysRevB.109.014210},
  urldate = {2026-03-03}
}

@article{Evers2008,
  title = {Anderson Transitions},
  author = {Evers, Ferdinand and Mirlin, Alexander D.},
  year = 2008,
  month = oct,
  journal = {Rev. Mod. Phys.},
  volume = {80},
  number = {4},
  pages = {1355--1417},
  publisher = {American Physical Society},
  doi = {10.1103/RevModPhys.80.1355},
  urldate = {2026-03-01}
}

@article{Ganeshan2015,
  title = {Nearest Neighbor Tight Binding Models with an Exact Mobility Edge in One Dimension},
  author = {Ganeshan, Sriram and Pixley, J. H. and Das Sarma, S.},
  year = 2015,
  month = apr,
  journal = {Phys. Rev. Lett.},
  volume = {114},
  number = {14},
  pages = {146601},
  publisher = {American Physical Society},
  doi = {10.1103/PhysRevLett.114.146601},
  urldate = {2026-03-02}
}

@article{Gautier2021,
  title = {Strongly Interacting Bosons in a Two-Dimensional Quasicrystal Lattice},
  author = {Gautier, Ronan and Yao, Hepeng and {Sanchez-Palencia}, Laurent},
  year = 2021,
  month = mar,
  journal = {Phys. Rev. Lett.},
  volume = {126},
  number = {11},
  pages = {110401},
  publisher = {American Physical Society},
  doi = {10.1103/PhysRevLett.126.110401},
  urldate = {2026-03-02}
}

@article{Goncalves2022,
  title = {Hidden Dualities in {{1D}} Quasiperiodic Lattice Models},
  author = {Gon{\c c}alves, Miguel and Amorim, Bruno and Castro, Eduardo and Ribeiro, Pedro},
  year = 2022,
  month = sep,
  journal = {SciPost Phys.},
  volume = {13},
  number = {3},
  pages = {046},
  issn = {2542-4653},
  doi = {10.21468/SciPostPhys.13.3.046},
  urldate = {2026-03-05}
}

@article{Goncalves2023,
  title = {Critical Phase Dualities in {{1D}} Exactly Solvable Quasiperiodic Models},
  author = {Gon{\c c}alves, Miguel and Amorim, Bruno and Castro, Eduardo V. and Ribeiro, Pedro},
  year = 2023,
  month = nov,
  journal = {Phys. Rev. Lett.},
  volume = {131},
  number = {18},
  pages = {186303},
  publisher = {American Physical Society},
  doi = {10.1103/PhysRevLett.131.186303},
  urldate = {2026-03-02}
}

@article{Goncalves2023a,
  title = {Renormalization Group Theory of One-Dimensional Quasiperiodic Lattice Models with Commensurate Approximants},
  author = {Gon{\c c}alves, Miguel and Amorim, B. and Castro, Eduardo V. and Ribeiro, Pedro},
  year = 2023,
  month = sep,
  journal = {Phys. Rev. B},
  volume = {108},
  number = {10},
  pages = {L100201},
  publisher = {American Physical Society},
  doi = {10.1103/PhysRevB.108.L100201},
  urldate = {2026-03-02}
}

@misc{Hou2026,
  title = {Nonlinear Photonic Tripartite Phase},
  author = {Hou, Xiangrui and Wang, Fangyu and Wu, Zhaoxin and Zhang, Shuming and Li, Shan-Zhong and Ying, Lei and Lin, Haiqing and Zhang, Baile and Li, Zhi and Zhu, Shi-Liang and Yang, Zhaoju},
  year = {2026},
  month = may,
  eprint = {2605.21983},
  archivePrefix = {arXiv},
  primaryClass = {physics.optics}
}

@article{Hu2025,
  title = {Hidden Self Duality and Exact Mobility Edges in Quasiperiodic Network Models},
  author = {Hu, Hai-Tao and Lin, Xiaoshui and Guo, Ai-Min and Guo, Guangcan and Lin, Zijing and Gong, Ming},
  year = 2025,
  month = jun,
  journal = {Phys. Rev. Lett.},
  volume = {134},
  number = {24},
  pages = {246301},
  publisher = {American Physical Society},
  doi = {10.1103/rl1f-ptzq},
  urldate = {2026-03-01}
}

@article{Huang2019,
  title = {Moir\'e Localization in Two-Dimensional Quasiperiodic Systems},
  author = {Huang, Biao and Liu, W. Vincent},
  year = 2019,
  month = oct,
  journal = {Phys. Rev. B},
  volume = {100},
  number = {14},
  pages = {144202},
  publisher = {American Physical Society},
  doi = {10.1103/PhysRevB.100.144202},
  urldate = {2026-03-04}
}

@article{Izrailev2012,
  title = {Anomalous Localization in Low-Dimensional Systems with Correlated Disorder},
  author = {Izrailev, F. M. and Krokhin, A. A. and Makarov, N. M.},
  year = 2012,
  month = mar,
  journal = {Phys. Rep.},
  volume = {512},
  number = {3},
  pages = {125--254},
  issn = {0370-1573},
  doi = {10.1016/j.physrep.2011.11.002},
  urldate = {2026-02-22}
}

@article{Jiang2019,
  title = {A {{Lieb-like}} Lattice in a Covalent-Organic Framework and Its {{Stoner}} Ferromagnetism},
  author = {Jiang, Wei and Huang, Huaqing and Liu, Feng},
  year = 2019,
  month = may,
  journal = {Nat. Commun.},
  volume = {10},
  number = {1},
  pages = {2207},
  publisher = {Nature Publishing Group},
  issn = {2041-1723},
  doi = {10.1038/s41467-019-10094-3},
  urldate = {2026-03-06}
}

@article{Jiang2024,
  title = {Exact Non-Hermitian Mobility Edges and Robust Flat Bands in Two-Dimensional {{Lieb}} Lattices with Imaginary Quasiperiodic Potentials},
  author = {Jiang, Xiang-Ping and Zeng, Weilei and Hu, Yayun and Liu, Peng},
  year = 2024,
  month = aug,
  journal = {New J. Phys.},
  volume = {26},
  number = {8},
  pages = {083020},
  publisher = {IOP Publishing},
  issn = {1367-2630},
  doi = {10.1088/1367-2630/ad6bb9},
  urldate = {2026-03-02}
}

@article{Kramer1993,
  title = {Localization: {{Theory}} and Experiment},
  shorttitle = {Localization},
  author = {Kramer, Bernhard and MacKinnon, Angus},
  year = 1993,
  month = dec,
  journal = {Rep. Prog. Phys.},
  volume = {56},
  pages = {1469--1564},
  doi = {10.1088/0034-4885/56/12/001}
}

@article{Lee1981,
  title = {Anderson Localization in Two Dimensions},
  author = {Lee, Patrick A. and Fisher, Daniel S.},
  year = 1981,
  month = sep,
  journal = {Phys. Rev. Lett.},
  volume = {47},
  number = {12},
  pages = {882--885},
  publisher = {American Physical Society},
  doi = {10.1103/PhysRevLett.47.882},
  urldate = {2026-02-22}
}

@article{Lee1985,
  title = {Disordered Electronic Systems},
  author = {Lee, Patrick A. and Ramakrishnan, T. V.},
  year = 1985,
  month = apr,
  journal = {Rev. Mod. Phys.},
  volume = {57},
  number = {2},
  pages = {287--337},
  publisher = {American Physical Society},
  doi = {10.1103/RevModPhys.57.287},
  urldate = {2026-03-01}
}

@article{Leykam2018,
  title = {Artificial Flat Band Systems: From Lattice Models to Experiments},
  shorttitle = {Artificial Flat Band Systems},
  author = {Leykam, Daniel and Andreanov, Alexei and Flach, Sergej},
  year = 2018,
  month = jan,
  journal = {Adv. Phys. X},
  volume = {3},
  number = {1},
  pages = {1473052},
  issn = {2374-6149},
  doi = {10.1080/23746149.2018.1473052},
  urldate = {2026-03-06}
}

@article{Li2016,
  title = {Quantum Nonergodicity and Fermion Localization in a System with a Single-Particle Mobility Edge},
  author = {Li, Xiaopeng and Pixley, J. H. and Deng, Dong-Ling and Ganeshan, Sriram and Das Sarma, S.},
  year = 2016,
  month = may,
  journal = {Phys. Rev. B},
  volume = {93},
  number = {18},
  pages = {184204},
  publisher = {American Physical Society},
  doi = {10.1103/PhysRevB.93.184204},
  urldate = {2026-03-05}
}

@article{Li2017,
  title = {Mobility Edges in One-Dimensional Bichromatic Incommensurate Potentials},
  author = {Li, Xiao and Li, Xiaopeng and Das Sarma, S.},
  year = 2017,
  month = aug,
  journal = {Phys. Rev. B},
  volume = {96},
  number = {8},
  pages = {085119},
  publisher = {American Physical Society},
  doi = {10.1103/PhysRevB.96.085119},
  urldate = {2026-03-05}
}

@article{Li2025a,
  title = {Anderson-Skin Dualism: {{A}} Boundary-Dependent Effect in Non-{{Hermitian}} Disordered Coupled Systems},
  shorttitle = {Anderson-Skin Dualism},
  author = {Li, Shan-Zhong and Li, Linhu and Zhu, Shi-Liang and Li, Zhi},
  year = 2025,
  month = nov,
  journal = {Phys. Rev. B},
  volume = {112},
  number = {20},
  pages = {L201108},
  publisher = {American Physical Society},
  doi = {10.1103/vz12-2qn3},
  urldate = {2026-07-25}
}

@article{Li2026,
  title = {Multifractal-Enriched Mobility Edges and Emergent Quantum Phases in {{Rydberg}} Atomic Arrays},
  author = {Li, Shan-Zhong and Zhang, Yi-Cai and Wang, Yucheng and Zhang, Shanchao and Zhu, Shi-Liang and Li, Zhi},
  year = 2026,
  journal = {Sci. China Phys. Mech. Astron.},
  volume = {69},
  number = {1},
  pages = {217212},
  issn = {1869-1927},
  doi = {10.1007/s11433-025-2774-2},
  urldate = {2026-07-14}
}

@article{Lieb1989,
  title = {Two Theorems on the {{Hubbard}} Model},
  author = {Lieb, Elliott H.},
  year = 1989,
  month = mar,
  journal = {Phys. Rev. Lett.},
  volume = {62},
  number = {10},
  pages = {1201--1204},
  publisher = {American Physical Society},
  doi = {10.1103/PhysRevLett.62.1201},
  urldate = {2026-03-06}
}

@article{Liu2020,
  title = {Localization, Phases, and Transitions in Three-Dimensional Extended {{Lieb}} Lattices},
  author = {Liu, Jie and Mao, Xiaoyu and Zhong, Jianxin and R{\"o}mer, Rudolf A.},
  year = 2020,
  month = nov,
  journal = {Phys. Rev. B},
  volume = {102},
  number = {17},
  pages = {174207},
  publisher = {American Physical Society},
  doi = {10.1103/PhysRevB.102.174207},
  urldate = {2026-03-02}
}

@article{MacKinnon1981,
  title = {One-Parameter Scaling of Localization Length and Conductance in Disordered Systems},
  author = {MacKinnon, A. and Kramer, B.},
  year = 1981,
  month = nov,
  journal = {Phys. Rev. Lett.},
  volume = {47},
  number = {21},
  pages = {1546--1549},
  publisher = {American Physical Society},
  doi = {10.1103/PhysRevLett.47.1546},
  urldate = {2026-03-01}
}

@article{Mott1987,
  title = {The Mobility Edge since 1967},
  author = {Mott, N.},
  year = 1987,
  month = jul,
  journal = {J. Phys. C Solid State Phys.},
  volume = {20},
  number = {21},
  pages = {3075},
  issn = {0022-3719},
  doi = {10.1088/0022-3719/20/21/008},
  urldate = {2026-03-04}
}

@article{Roati2008,
  title = {Anderson Localization of a Non-Interacting {{Bose}}--{{Einstein}} Condensate},
  author = {Roati, Giacomo and D'Errico, Chiara and Fallani, Leonardo and Fattori, Marco and Fort, Chiara and Zaccanti, Matteo and Modugno, Giovanni and Modugno, Michele and Inguscio, Massimo},
  year = 2008,
  month = jun,
  journal = {Nature},
  volume = {453},
  number = {7197},
  pages = {895--898},
  publisher = {Nature Publishing Group},
  issn = {1476-4687},
  doi = {10.1038/nature07071},
  urldate = {2026-02-22}
}

@article{Roche1997,
  title = {Conductivity of Quasiperiodic Systems: A Numerical Study},
  shorttitle = {Conductivity of Quasiperiodic Systems},
  author = {Roche, S. and Mayou, D.},
  year = 1997,
  month = sep,
  journal = {Phys. Rev. Lett.},
  volume = {79},
  number = {13},
  pages = {2518--2521},
  publisher = {American Physical Society},
  doi = {10.1103/PhysRevLett.79.2518},
  urldate = {2026-03-04}
}

@article{Rossignolo2019,
  title = {Localization Transitions and Mobility Edges in Coupled {{Aubry-Andr\'e}} Chains},
  author = {Rossignolo, M. and Dell'Anna, L.},
  year = 2019,
  month = feb,
  journal = {Phys. Rev. B},
  volume = {99},
  number = {5},
  pages = {054211},
  publisher = {American Physical Society},
  doi = {10.1103/PhysRevB.99.054211},
  urldate = {2026-03-02}
}

@article{Sarker1981,
  title = {Scaling Theory of {{Anderson}} Localization: A Renormalization-Group Approach},
  shorttitle = {Scaling Theory of Anderson Localization},
  author = {Sarker, Sanjoy and Domany, Eytan},
  year = 1981,
  month = jun,
  journal = {Phys. Rev. B},
  volume = {23},
  number = {11},
  pages = {6018--6036},
  publisher = {American Physical Society},
  doi = {10.1103/PhysRevB.23.6018},
  urldate = {2026-03-01}
}

@article{Schwartz2007,
  title = {Transport and {{Anderson}} Localization in Disordered Two-Dimensional Photonic Lattices},
  author = {Schwartz, Tal and Bartal, Guy and Fishman, Shmuel and Segev, Mordechai},
  year = 2007,
  month = mar,
  journal = {Nature},
  volume = {446},
  number = {7131},
  pages = {52--55},
  publisher = {Nature Publishing Group},
  issn = {1476-4687},
  doi = {10.1038/nature05623},
  urldate = {2026-02-22}
}

@article{SergeAubry1980,
  title = {Analyticity breaking and Anderson localization in incommensurate lattices},
  author = {{Serge Aubry} and {Gilles Andr\'e}},
  year = 1980,
  journal = {Ann. Israel Phys. Soc.},
  volume = {3},
  pages = {133--140}
}

@article{Strkalj2022,
  title = {Coexistence of Localization and Transport in Many-Body Two-Dimensional {{Aubry-Andr\'e}} Models},
  author = {{\v S}trkalj, Antonio and Doggen, Elmer V. H. and Castelnovo, Claudio},
  year = 2022,
  month = nov,
  journal = {Phys. Rev. B},
  volume = {106},
  number = {18},
  pages = {184209},
  publisher = {American Physical Society},
  doi = {10.1103/PhysRevB.106.184209},
  urldate = {2026-03-02}
}

@article{Sutherland1986,
  title = {Localization of Electronic Wave Functions Due to Local Topology},
  author = {Sutherland, Bill},
  year = 1986,
  month = oct,
  journal = {Phys. Rev. B},
  volume = {34},
  number = {8},
  pages = {5208--5211},
  publisher = {American Physical Society},
  doi = {10.1103/PhysRevB.34.5208},
  urldate = {2026-07-20}
}

@article{Szabo2020,
  title = {Mixed Spectra and Partially Extended States in a Two-Dimensional Quasiperiodic Model},
  author = {Szab{\'o}, Attila and Schneider, Ulrich},
  year = 2020,
  month = jan,
  journal = {Phys. Rev. B},
  volume = {101},
  number = {1},
  pages = {014205},
  publisher = {American Physical Society},
  doi = {10.1103/PhysRevB.101.014205},
  urldate = {2026-03-02}
}

@article{Thouless1974,
  title = {Electrons in Disordered Systems and the Theory of Localization},
  author = {Thouless, D. J.},
  year = 1974,
  month = oct,
  journal = {Phys. Rep.},
  volume = {13},
  number = {3},
  pages = {93--142},
  issn = {0370-1573},
  doi = {10.1016/0370-1573(74)90029-5},
  urldate = {2026-02-22}
}

@article{Wang2020,
  title = {One-Dimensional Quasiperiodic Mosaic Lattice with Exact Mobility Edges},
  author = {Wang, Yucheng and Xia, Xu and Zhang, Long and Yao, Hepeng and Chen, Shu and You, Jiangong and Zhou, Qi and Liu, Xiong-Jun},
  year = 2020,
  month = nov,
  journal = {Phys. Rev. Lett.},
  volume = {125},
  number = {19},
  pages = {196604},
  publisher = {American Physical Society},
  doi = {10.1103/PhysRevLett.125.196604},
  urldate = {2026-03-02}
}

@article{Wang2021,
  title = {Duality between Two Generalized {{Aubry-Andr\'e}} Models with Exact Mobility Edges},
  author = {Wang, Yucheng and Xia, Xu and Wang, Yongjian and Zheng, Zuohuan and Liu, Xiong-Jun},
  year = 2021,
  month = may,
  journal = {Phys. Rev. B},
  volume = {103},
  number = {17},
  pages = {174205},
  publisher = {American Physical Society},
  doi = {10.1103/PhysRevB.103.174205},
  urldate = {2026-03-02}
}

@article{Wang2022,
  title = {Quantum Phase with Coexisting Localized, Extended, and Critical Zones},
  author = {Wang, Yucheng and Zhang, Long and Sun, Wei and Poon, Ting-Fung Jeffrey and Liu, Xiong-Jun},
  year = 2022,
  month = oct,
  journal = {Phys. Rev. B},
  volume = {106},
  number = {14},
  pages = {L140203},
  publisher = {American Physical Society},
  doi = {10.1103/PhysRevB.106.L140203},
  urldate = {2026-03-05}
}

@article{Wang2023,
  title = {Two-Dimensional Vertex-Decorated {{Lieb}} Lattice with Exact Mobility Edges and Robust Flat Bands},
  author = {Wang, Yucheng and Zhang, Long and Wan, Yuhao and He, Yu and Wang, Yongjian},
  year = 2023,
  month = apr,
  journal = {Phys. Rev. B},
  volume = {107},
  number = {14},
  pages = {L140201},
  publisher = {American Physical Society},
  doi = {10.1103/PhysRevB.107.L140201},
  urldate = {2026-03-01}
}

@article{Wang2023a,
  title = {Engineering Mobility in Quasiperiodic Lattices with Exact Mobility Edges},
  author = {Wang, Zhenbo and Zhang, Yu and Wang, Li and Chen, Shu},
  year = 2023,
  month = nov,
  journal = {Phys. Rev. B},
  volume = {108},
  number = {17},
  pages = {174202},
  publisher = {American Physical Society},
  doi = {10.1103/PhysRevB.108.174202},
  urldate = {2026-03-05}
}

@article{Wang2023b,
  title = {Numerical Investigation of Localization in Two-Dimensional Quasiperiodic Mosaic Lattice},
  author = {Wang, Hui-Hui and Wang, Si-Si and Yu, Yan and Zhang, Biao and Dai, Yi-Ming and Chen, Hao-Can and Zhang, Yi-Cai and Zhang, Yan-Yang},
  year = 2023,
  month = feb,
  journal = {J. Phys. Condens. Matter},
  volume = {35},
  number = {13},
  pages = {135301},
  publisher = {IOP Publishing},
  issn = {0953-8984},
  doi = {10.1088/1361-648X/acb67c},
  urldate = {2026-03-04}
}

@article{Xu2021,
  title = {Exact Mobility Edges and Topological Phase Transition in Two-Dimensional Non-Hermitian Quasicrystals},
  author = {Xu, Zhi-Hao and Xia, Xu and Chen, Shu},
  year = 2021,
  month = dec,
  journal = {Sci. China Phys. Mech. Astron.},
  volume = {65},
  number = {2},
  pages = {227211},
  issn = {1869-1927},
  doi = {10.1007/s11433-021-1802-4},
  urldate = {2026-03-02}
}

@article{Yang2024,
  title = {Exploring Multifractal Critical Phases in Two-Dimensional Quasiperiodic Systems},
  author = {Yang, Chao and Yang, Weizhe and Wang, Yongjian and Wang, Yucheng},
  year = 2024,
  month = oct,
  journal = {Phys. Rev. A},
  volume = {110},
  number = {4},
  pages = {042205},
  publisher = {American Physical Society},
  doi = {10.1103/PhysRevA.110.042205},
  urldate = {2026-03-02}
}

@article{Yao2019,
  title = {Critical Behavior and Fractality in Shallow One-Dimensional Quasiperiodic Potentials},
  author = {Yao, Hepeng and Khoudli, Alice and Bresque, L{\'e}a and {Sanchez-Palencia}, Laurent},
  year = 2019,
  month = aug,
  journal = {Phys. Rev. Lett.},
  volume = {123},
  number = {7},
  pages = {070405},
  publisher = {American Physical Society},
  doi = {10.1103/PhysRevLett.123.070405},
  urldate = {2026-03-05}
}

@article{Yuan2000,
  title = {Energy Spectra, Wave Functions, and Quantum Diffusion for Quasiperiodic Systems},
  author = {Yuan, H. Q. and Grimm, U. and Repetowicz, P. and Schreiber, M.},
  year = 2000,
  month = dec,
  journal = {Phys. Rev. B},
  volume = {62},
  number = {23},
  pages = {15569--15578},
  publisher = {American Physical Society},
  doi = {10.1103/PhysRevB.62.15569},
  urldate = {2026-03-02}
}

@article{Gao2025a,
  title = {Floquet Engineering of Mobility Edges in Quasiperiodic Mosaic Lattice},
  author = {Gao, Xin and Zhou, Zi-Hang and Cao, Ji and Cui, Wen-Xue and Zhang, Shou and Wang, Hong-Fu},
  journal = {Adv. Quantum Technol.},
  volume = {8},
  pages = {e00287},
  year = {2025},
  doi = {10.1002/qute.202500287}
}

@article{Zhou2021,
  title = {Floquet engineering of topological localization transitions and mobility edges in one-dimensional non-{{Hermitian}} quasicrystals},
  author = {Zhou, Longwen},
  journal = {Phys. Rev. Res.},
  volume = {3},
  issue = {3},
  pages = {033184},
  numpages = {19},
  year = {2021},
  month = {Aug},
  publisher = {American Physical Society},
  doi = {10.1103/PhysRevResearch.3.033184}
}

@article{Lin2026,
  title = {Topological {{Anderson}} insulators by latent symmetry},
  author = {Lin, Jing-Run and Wang, Shuo and Li, Hui and Zuo, Zheng-Wei},
  journal = {Phys. Rev. B},
  volume = {113},
  issue = {9},
  pages = {094201},
  year = {2026},
  month = {Mar},
  publisher = {American Physical Society},
  doi = {10.1103/PhysRevB.113.094201}
}

@article{Zhang2025,
  title = {Observation of higher-order time-dislocation topological modes},
  author = {Zhang, Jia-Hui and Mei, Feng and Li, Y. and Lee, C. and Ma, J. and Xiao, Liantuan and Jia, Suotang},
  journal = {Nat. Commun.},
  volume = {16},
  pages = {2050},
  year = {2025},
  doi = {10.1038/s41467-025-56717-w},
  url = {https://doi.org/10.1038/s41467-025-56717-w}
}

@article{Zhang2024,
  title = {Dynamical Detection of Topological Spectral Density},
  author = {Zhang, Jia-Hui and Mei, Feng and Xiao, Liantuan and Jia, Suotang},
  journal = {Phys. Rev. Lett.},
  volume = {132},
  pages = {036603},
  year = {2024},
  doi = {10.1103/PhysRevLett.132.036603},
  url = {https://doi.org/10.1103/PhysRevLett.132.036603}
}

@article{Mei2020,
  title = {Digital Simulation of Topological Matter on Programmable Quantum Processors},
  author = {Mei, Feng and Guo, Qihao and Yu, Ya-Fei and Xiao, Liantuan and Zhu, Shi-Liang and Jia, Suotang},
  journal = {Phys. Rev. Lett.},
  volume = {125},
  pages = {160503},
  year = {2020},
  doi = {10.1103/PhysRevLett.125.160503},
  url = {https://doi.org/10.1103/PhysRevLett.125.160503}
}

@article{Qian2025,
  title = {Programmable higher-order nonequilibrium topological phases on a superconducting quantum processor},
  author = {Qian, H. and Gong, M. and Zhang, J. and Guo, S. and Zha, C. and Chen, F. and Ye, Y. and Wu, Y. and Cao, S. and Ying, C. and Zhu, Q. and Huang, H. L. and Zhao, Y. and Li, S. W. and Yu, J. and Fan, D. and Wu, D. and Su, H. and Deng, H. and Rong, H. and Li, Y. and Zhang, K. and Chung, T. H. and Liang, F. and Lin, J. and Xu, Y. and Guo, C. and Li, N. and Yan, K. and Su, F. F. and Wu, G. and Huo, Y. H. and Peng, Cheng-Zhi and Lu, Chao-Yang and Mei, Feng and Jia, Suotang and Zhu, Xiaobo and Pan, Jian-Wei},
  journal = {Science},
  volume = {390},
  pages = {930},
  year = {2025},
  doi = {10.1126/science.adp6802},
  url = {https://doi.org/10.1126/science.adp6802}
}

\end{document}